# A Probabilistic Domain-knowledge Framework for Nosocomial Infection Risk Estimation of Communicable Viral Diseases in Healthcare Personnel: A Case Study for COVID-19

Phat K. Huynh, *Member, IEEE*, Arveity R. Setty, Om P. Yadav, *Fellow, IEEE,* and Trung Q. Le, *Fellow*, *IEEE*

*Abstract*—Hospital-acquired infections of communicable viral diseases (CVDs) are posing a tremendous challenge to healthcare workers globally. Healthcare personnel (HCP) is facing a consistent risk of hospital-acquired infections, and subsequently higher rates of morbidity and mortality. We proposed a domain-knowledge-driven infection risk model to quantify the individual HCP and the population-level healthcare facility risks. For individual-level risk estimation, a time-variant infection risk model is proposed to capture the transmission dynamics of CVDs. At the population-level, the infection risk is estimated using a Bayesian network model constructed from three feature sets, including individual-level factors, engineering control factors, and administrative control factors. The sensitivity analyses indicated that the uncertainty in the individual infection risk can be attributed to two variables: the number of close contacts and the viral transmission probability. The model validation was implemented in the transmission probability model, individual-level risk model, and population-level risk model using a Coronavirus disease case study. Regarding the first, multivariate logistic regression was applied for a cross-sectional data in the UK with an AIC value of 7317.70 and a 10-fold cross validation accuracy of 78.23%. For the second model, we collected laboratory-confirmed COVID-19 cases of HCP in different occupations. The occupation-specific risk evaluation suggested the highest-risk occupations were registered nurses, medical assistants, and respiratory therapists, with estimated risks of 0.0189, 0.0188, and 0.0176, respectively. To validate the population-level risk model, the infection risk in Texas and California was estimated. The proposed model will significantly influence the PPE allocation and safety plans for HCP.

## I. INTRODUCTION

Nosocomial infections (i.e., hospital-acquired infections) of communicable viral diseases (CVDs) (e.g., influenza virus, hepatitis A virus, and rotavirus infections) have posed huge challenges to public health organizations and the functioning of healthcare systems globally [1], especially during the Coronavirus disease (COVID-19) outbreak in 2019. Hospitals saw an increasing number of outbreaks of CVDs over the last decade, which had negative impacts on patient and healthcare workers' morbidity and mortality [2, 3]. Among nosocomial infections, healthcare personnel (HCP) experience the highest risk [4, 5] because of the direct or indirect contact with infected patients and virus-contaminated surfaces. Subsequently, these workers may spread the virus to non-infectious patients, coworkers, and their family members. In addition, containment and preventive measures in hospital settings usually overlook asymptomatic individuals and "super spreader" events [6, 7]. Therefore, mitigating and preventing nosocomial infections in hospitals is an urgent and important task to lower the risk of contracting CVDs for HCP.

Modeling of nosocomial HCP infections in hospitals has been based on mathematical models to qualitatively capture the dynamics of CVDs and the effects of different control measures [8, 9] when only limited data are available. One traditional mathematical model of disease spread is the compartmental SEIR (Susceptible-Exposed-Infected-Recovered) model [10]. It divides a population into four different compartments or sub-groups (susceptible, exposed, infected, and recovered individuals) and employs deterministic ordinary differential equations to model the spread of a CVD. In the literature, there are many variants of this model (e.g., SIS, SIRD, MSIR, and MSEIR model). These models consider the population as homogenous without individual interactions (e.g., patients and HCP); therefore, they fail to capture the individual contact process and the effects of individual risk and protective factors [11]. To overcome the limitations of the classic models, complex systems approaches using cellular automata (CA) theory have been proposed to model location-specific dynamics of susceptible populations and the probabilistic nature of disease transmission [12, 13]. The major drawback of CA models is its insufficiency in characterizing the spatial temporal information of individuals' movements and interactions [14]. Agent-based modeling (ABM) was proposed to address the limitations of CA models by accounting for the movement of individual disease carriers and the contact network of people

This work was supported in National Institute of General Medical Sciences of the National Institutes of Health (NIH) under Award Number U54GM128729.

Phat K. Huynh, PhD student, is with North Dakota State University, ND 58108 USA (e-mail: phat.huynh@ndsu.edu).

Arveity R. Setty, MD, is with University of North Dakota, ND 58202 USA. He is working in Sanford hospital, ND 58102 USA (e-mail: Arveity.Setty@SanfordHealth.org).

Om P. Yadav, PhD, is with North Carolina A&T State University, NC 27411 USA. (e-mail: oyadav@ncat.edu).

*Corresponding author: Trung. Q. Le, PhD, is with North Dakota State University, ND 58102 (e-mail: trung.q.le@ndsu.edu).

[15]. Although the ABM approach can capture the spread of a CVD in a spatial region (e.g., hospital) over time and estimate the risk of viral infection, it requires a large amount of information of individuals' movement and high computational cost. Moreover, individuals' movements are highly restricted in hospital settings, especially for patients who have positive test results for infectious diseases.

Quantitative models have also been used as an alternative to mathematical models to quantify the effects of protective or risk factors on the infection risk of HCP over time. These models capture the disease transmission dynamics within the hospital, HCP-related risk factors of infection, and other patients and HCP as sources of infection [16]. Here, variables are treated as time-dependent variables. Two classes of quantitative models, namely measure of association and statistical survival analysis, have been proposed to estimate HCP infection risk. The measure of association approaches quantifies the relationship between the exposed and diseased HCP groups by using the adjusted odds ratio (aOR), risk difference (RD), and relative risk (RR) as the risk measures [5, 17-20]. To capture the changes of HCP's characteristics and infection risk over time, survival analysis models are used to estimate the HCP infection risk and the expected duration of time until a viral infection occurs [21, 22]. Although time-dependent variables have been considered in the survival analysis models, the stochastic nature of epidemiological dynamics and individual interactions have not been investigated. Estimating the HCP infection risk of nosocomial infection is important to answer epidemiological questions in the hospital settings and provide information for PPE allocation, safety plans for HCP, and staffing strategies.

To overcome the above research gaps, this paper proposes a probabilistic domain-knowledge model of the infection risk of CVDs for HCP. The proposed model was formulated for the infection risk estimation at both individual and population levels with respect to three modes of transmissions: 1) direct contact of susceptible HCP with other infectious individuals including patients and coworkers, 2) airborne viruses, and 3) contaminated equipment and surfaces. The individual-level risk model was built based on the population grouping in the SEIR model with the consideration of the time-varying confounders to capture the dynamical contagious disease transmission mechanism. At the population-level, three subsets of features, which are introduced in Sub-section II.B, were constructed and represented by a Bayesian network [23], from which the probability of transmission from patients to HCP was estimated. The main contributions of this paper are 1) a novel time-variant infection risk analysis model to characterize the dynamics of the disease exposure risk in HCP over time and 2) an individual-specific and domain-knowledge driven infection risk to quantify the complexities of HCP's infection risk. The remainder of the manuscript is organized as follows: Section II elaborates the proposed model, model formulation, and validation; the results with sensitivity analysis and the case study on the COVID-19 are presented in Section III, discussion and conclusions are provided in Sections IV and V.

## II. METHODOLOGIES

The proposed framework consists of two sub-models: (1) an individual-level infection risk model that quantifies the risk of infection of an HCP, and (2) a population-level infection risk indicator model that estimates the infection risk under working conditions at a medical facility. The output from the first sub-model serves as an input for the estimation of the population infection risk in the second model. Other inputs, such as engineering control and administrative factors, were also considered in the estimation of population risk.

### A. Individual-level infection risk model

The individual infection risk model aims to quantify the potential risk of infection associated with a healthcare worker subject to nosocomial infection, whose job functions require working in proximity of patients. The proposed individual-level infection risk model is formulated using the population grouping approach in the compartmental SEIR model [10], in which the population is divided into different compartments (i.e., Susceptible ($S$), Exposed ($E$), Infectious ($I$), or Recovered ($R$)). However, susceptible ($S$) and recovered individuals ($R$) cannot transmit the virus during the length of a hospital stay, hence we do not consider these compartments in our model. Moreover, we do not assume that the recovered patients confer immunity to reinfection when being released from isolation. HCP coworkers have also been shown to contribute significantly to virus spread within the healthcare setting if contracting a virus [21, 24]. To capture the virus transmission mechanism, the healthcare worker group ($HW$) is added to model the HCP-HCP transmission, and the infectious individuals are further classified into two sub-groups: the infection-confirmed group ($IC$) and the infection-suspected ($IS$) group. Infection-confirmed individuals are those who have lab-confirmed infections *(e.g.,* individuals have tested positive for COVID-19 using the polymerase chain reaction (PCR) test), and the infection-suspected group includes individuals who are suspected to have the virus infection because they developed symptoms but have never tested for the infectious disease. In total, four groups ($E, IC, IS, HW$) are considered to model the individual HCP infection risk. We denote the potential infection risk of the HCP $j$ at location $i$ (e.g., hospitals) over time from $t_1$ to $t_2$ as $PIR_{i,j}^{(t_1:t_2)}$, which is the cumulative risk of viral infection after contacting patients and contaminated surfaces. We denote $N_{E,j}^{(t_1:t_2)}, N_{IC,j}^{(t_1:t_2)}, N_{IS,j}^{(t_1:t_2)}$, and $N_{HW,j}^{(t_1:t_2)}$ as the number of exposed cases, infection-confirmed, infection-suspected, and colleagues that an HCP $j$ has contacted with over the time $(t_1:t_2)$, which is contracted as $(\cdot)$ (e.g., $N_{E,j}^{(t_1:t_2)} = N_{E,j}^{(\cdot)}$).

An HCP $j$ is assumed to have $CC_{X,k}^{(t_1:t_2)}$ independent close contacts with an individual $k$. Next, we denote $p_{X,k \to j}^{(\cdot)}$ as the probability of viral transmission from individual $k$ to the HCP $j$, with $X \in \{E, IC, IS, HW\}$ being the compartment indicator of person $k$. Here, if the probability $p_{X,k \to j}^{(\cdot)}$ is constant, the viral transmission mechanism is modelled as a binomial process $Bin(CC_{X,k}^{(\cdot)}, p_{X,k \to j}^{(\cdot)})$ [25], and there are $N_{E,j}^{(\cdot)} + N_{IC,j}^{(\cdot)} + N_{IS,j}^{(\cdot)} + N_{HW,j}^{(\cdot)}$ binomial processes in total. The sequence of contacts of HCP $j$ ordered by time will be superscripted by person index $k(m)$ and compartment index $X(m)$ as follows:

$$\boldsymbol{C}^{(t_1:t_2)} = \left\{ C_m^{k(m),X(m)} \middle| k(m) = 1, \dots, N_{X(m),j}^{(\cdot)} \right\} \quad (1)$$

where $X(m) = \{E, IC, IS, HW\}$, $m$ is the temporal order of close contacts from which the HCP $j$ contracts the virus, $C_m^{k(m),X(m)} = 1$ if the HCP $j$ contracts the virus at the $m^{th}$ close contact, $C_m^{k(m),X(m)} = 0$ otherwise. As a result, the risk $PIR_{i,j}^{(\cdot)}$, is estimated as:

$$PIR_{i,j}^{(\cdot)} = \sum_{m=1}^{|C^{(\cdot)}|} P\left(C_m^{k(m),X(m)} = 1, \boldsymbol{C}_{1:m-1}^{k(m),X(m)} = \boldsymbol{0}\right) \quad (2)$$

where $|\boldsymbol{C}^{(\cdot)}|$ is the total number of contacts and $\boldsymbol{C}_{1:m-1}^{k(m),X(m)} = \boldsymbol{0}$ means all previous $m - 1$ contacts are the failed transmissions. Given the assumption of independent close contacts, (2) can be expressed as:

$$PIR_{i,j}^{(\cdot)} = \sum_{m=1}^{|C^{(\cdot)}|} \left[\prod_{r=1}^{m-1}\left(1 - p_{X(r),k(r)\to j}^{(\cdot)}\right)\right] p_{X(m),k(m)\to j}^{(\cdot)} \quad (3)$$

If we denote $TP_-^{j,k}$ and $TP_+^{j,k}$ as the impatient admission time and the recovery time of an individual $k$ with whom the HCP $j$ has close contacts, the time interval $[TP_-^{j,k}, TP_+^{j,k}]$ is the virus exposure period for the HCP $j$ with the person $k$. Therefore, $p_{X(r),k(r)\to j}^{(t_1:t_2)}$ can be reduced to $p_{X(r),k(r)\to j}^{\left(\max\{t_1,TP_-^{j,k(r)}\}:\min\{t_2,TP_+^{j,k(r)}\}\right)}$. If $p_{X,k\to j}^{(\cdot)}$ varies over time, the constant $p_{X,k\to j}^{(\cdot)}$ assumption is relaxed by considering the cumulative distribution function that describes the probability of infection up to time $t$: $F(t) = P(T \le t) = 1 - \exp\left(-\int_0^t h(t)dt\right)$, in which $T$ is the infection time and $h(t)$ is the hazard function. We assume $h(t) = 0$ over the time of no close contacts. Hence, $PIR_{i,j}^{(\cdot)} = P(t_1 \le T \le t_2)$ is estimated as:

$$\begin{aligned} PIR_{i,j}^{(\cdot)} &= 1 - \exp\left(-\int_0^{t_2} h(t)dt\right) - \exp\left(-\int_0^{t_1} h(t)dt\right) \\ &= \sum_{m=1}^{|C^{(\cdot)}|}\left\{1 - \exp\left(-\int_0^{\tau_m} h_m(t)dt\right)\right\} \end{aligned} \quad (4)$$

where $\tau_r$ is the length of the $m^{th}$ close contact with person $k(m)$, and $h_m(t)$ is the cumulative infection time distribution function for the $m^{th}$ close contact. The probability $p_{X(r),k(r)\to j}^{(\cdot)}$ and $h_m(t)$ depend on various factors including HCP-dependent features, patient-dependent features, patient-HCP interactions, HCP-HCP interactions, and healthcare facilities' conditions. A logistic regression model is further established to estimate the probability $p_{X(r),k(r)\to j}^{(\cdot)}$ as:

$$\log\left[\frac{p_{X(r),k(r)\to j}^{(\cdot)}}{1 - p_{X(r),k(r)\to j}^{(\cdot)}}\right] = \boldsymbol{Z}^T\boldsymbol{\beta} \quad (5)$$

$$p_{X(r),k(r)\to j}^{(\cdot)} = P\left(Y_{X(r),k(r)\to j}^{(\cdot)} = 1\right) = \frac{\exp(\boldsymbol{Z}^T\boldsymbol{\beta})}{1 + \exp(\boldsymbol{Z}^T\boldsymbol{\beta})} \quad (6)$$

where $Y_{X(r),k(r)\to j}^{(\cdot)}$ is the indicator variable ($Y_{X(r),k(r)\to j}^{(\cdot)} = 1$ means that HCP $j$ has contracted the virus via the contact with person $k(r)$ and $Y_{X(r),k(r)\to j}^{(\cdot)} = 0$ if HCP $j$ has failed to contract the virus), $\boldsymbol{Z}$ is the covariate vector including the factors influencing the response and $\boldsymbol{\beta}$ is the coefficient vector.

*B. Population risk indicator model*

The population risk indicator quantifies the potential viral infection risk associated with a hospital/clinic over the time period $[t_1: t_2]$. The population risk, annotated as $PIR_i^{(t_1:t_2)}$, is interpreted as the probability that an HCP contracts the disease under working conditions at place $i$ given the information about the individual-level infection risk of all HCP at place $i$ and the external factors. At this level, external factors from engineering and administrative controls within the hospital are considered. Those are the factors that affect the population-level infection risk apart from the individual-level risk. Representative examples of engineering controls are high-efficiency air, ventilation rates at the workplace, and infection isolation rooms for aerosol generating procedures. Administrative controls include formal HCP training regarding protective personal equipment (PPE), training on risk factors and resources to promote personal hygiene. The $PIR_i^{(t_1:t_2)}$ is computed using logistic function as:

$$PIR_i^{(\cdot)} = \left\{1 + \exp\left[-\sum_j \frac{f\left(\boldsymbol{PIR}_{i,j=1,\ldots n_{HCP}}^{(\cdot)}, \boldsymbol{F}\right)}{\tau}\right]\right\}^{-1} \quad (7)$$

where $\boldsymbol{PIR}_{i,j=1,\ldots n_{HCP}}^{(\cdot)} = \left[PIR_{i,1}^{(\cdot)}, PIR_{i,2}^{(\cdot)}, \ldots PIR_{i,n_{HCP}}^{(\cdot)}\right]^T$ is the vector of individual infection risk estimates of a total number of $n_{HCP}$ HCP, $\tau$ is the scaling parameter, $\boldsymbol{F} = \{F_i\}$ is the vector of engineering control and administrative control factors. We denote $f(\cdot)$ as the abbreviated notation for the function of $PIR_{i,j}^{(\cdot)}$ and $\boldsymbol{F}$ in (9). The function $f(\cdot)$ can be simply formulated as a linear regression model such that:

$$f(\cdot) = \boldsymbol{\alpha}\boldsymbol{PIR}_{i,j=1,\ldots n_{HCP}}^{(\cdot)} + w_1 F_1 + \cdots + w_n F_n + b \quad (8)$$

where $\boldsymbol{\alpha}, w_i$, and $b$ are the model parameters. Alternatively, the population risk $PER_i^{(\cdot)}$ is estimated using a Bayesian network when we have access to the domain knowledge that describe the relationships between the control factors and the infection risk at population level and individual level. Here, the Bayesian network model [26] is employed to incorporate the domain knowledge that influences the virus spread. The network is formulated based on three subsets of factors from the literature that affect the risk of infection including 1) individual-level factors, 2) engineering control factors, and 3) administrative control factors (see Fig. 1). Individual-level factors include patient characteristics (e.g., time from exposure to symptom onset) clinical severity of patients), HCP-dependent factors (e.g., PPE sufficiency level, close contacts with patients, exposure level to infection, working hours per week), and intervention-related risks (e.g., endotracheal intubation, high flow nasal canula (HFNC). External factors consist of engineering control factors (e.g., high-efficiency air, ventilation rates, airborne infection isolation rooms) and administrative control factors (e.g., formal HCP training on PPE and disease risk factors, resources to promote personal hygiene). These factors are annotated as $\boldsymbol{ILF}$, $\boldsymbol{ECF}$, and $\boldsymbol{ACF}$ respectively. Hence, using the chain rule of the Bayesian network [27], the risk $PIR_i^{(\cdot)}$ is estimated as:

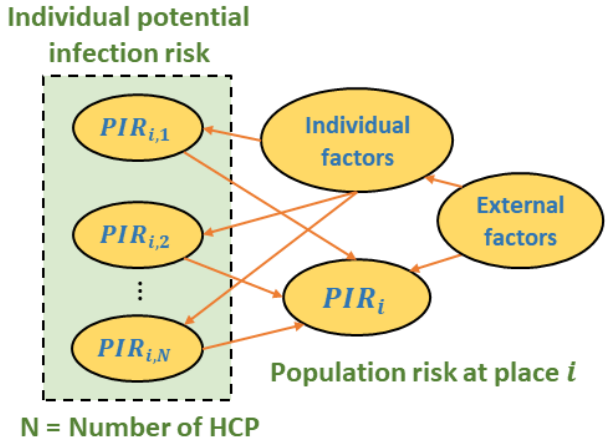

**Fig. 1.** Directed acyclic graph (DAG) representation of the protective and risk factors influencing the risk of infection of HCP, where the directed edges between two nodes of the Bayesian network indicate the conditional probabilistic dependencies.

$$PIR_i^{(\cdot)} = P\left(X_{PIR_i^{(\cdot)}} = 1 \middle| PIR_{i,j}^{(\cdot)}, ECF, ACF, ILF\right)$$

$$= \frac{P\left(X_{PIR_i^{(\cdot)}}, PIR_{i,j}^{(\cdot)}, ECF, ACF, ILF\right)}{P(ECF)P(ACF)P(ILF|ECF, ACF)P\left(PIR_{i,j}^{(\cdot)} \middle| ILF\right)} \quad (9)$$

where $P(\cdot)$ is the probability function, and $X_{PIR_i^{(\cdot)}}$ is the indicator variable ($X_{PIR_{i,j}^{(\cdot)}} = 1$ indicates that an HCP contracts the disease and $X_{PIR_{i,j}^{(\cdot)}} = 0$ if they do not).

## III. IMPLEMENTATION AND RESULTS

### A. Sensitivity analysis using simulated data

Variance-based sensitivity analysis was utilized to investigate the uncertainty of HCP's potential infection risk output caused by the variance of the input variables.

*1) The measure of sensitivity of $PIR_{i,j}^{(\cdot)}$ to $p_{X(m),k(m)\to j}^{(\cdot)}$ and close contact sequence*

The dependence of the potential infection risk on the probability of viral transmission and close contact sequence for an HCP was analyzed. For each close contact, the probability $p_{X(m),k(m)\to j}^{(\cdot)}$ has 3 levels: $P_{low} = 0.01, P_{medium} = 0.05$, and $P_{high} = 0.1$. $PIR_{i,j}^{(\cdot)}$'s for different numbers of close contacts $|C^{(\cdot)}|$ were estimated by (3). For illustration, the results for $|C^{(\cdot)}| = 2$ and $|C^{(\cdot)}| = 3$ are shown in Fig. 2. According to the results, the mean level ($\pm$ SD) of $PIR_{i,j}^{(\cdot)}$ for $|C^{(\cdot)}| = 2$ was $0.1038 \pm 0.0523$, which was lower than that for $|C^{(\cdot)}| = 3$ at $0.1516 \pm 0.0583$. The mean value of the individual risk escalated together with the standard deviation values as the number of contacts increased. In addition, the estimated $PIR_{i,j}^{(\cdot)}$ was not influenced by the time order of the close contacts, e.g., the same $PIR_{i,j}^{(\cdot)} = 0.1065$ for three sequences: 011, 101, 110, where 0 and 1 are the encoded values for $P(Low)$ and $P(medium)$ respectively. The results are from the assumption of temporal independence between close contacts However, the risk would increase when the probability $p_{X(m),k(m)\to j}^{(\cdot)}$ for each contact raised to a higher value, hence the probabilities collectively contributed to the value of risk.

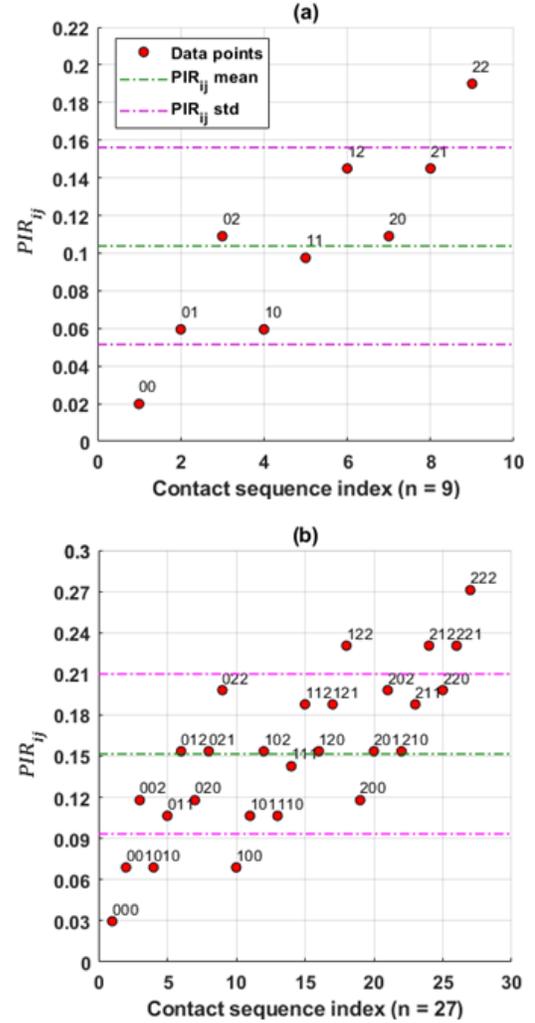

**Fig. 2.** Sensitivity analysis of the impact of probability of viral transmission and the number of close contacts on $PIR_{i,j}^{(t_1:t_2)}$. We estimated values of $PIR_{i,j}^{(t_1:t_2)}$ for the synthesized data with three levels of $p_{X(m),k(m)\to j}^{(\cdot)}$: $P_{low} = 0.01, P_{medium} = 0.05, P_{high} = 0.1$. Panel (a): the estimated $PIR_{i,j}^{(t_1:t_2)}$ for $|C^{(\cdot)}| = 2$, i.e., two close contacts; therefore, there are $n = 3^2 = 9$ possible contact sequences with different combinations of $p_{X(m),k(m)\to j}^{(\cdot)}$ levels, and those combinations are encoded in the form $X_1 X_2 \ldots X_n$, where $X_1, X_2, \ldots, X_n \in \{0,1,2\}$, which corresponds to low, medium, and high level of $p_{X(m),k(m)\to j}^{(\cdot)}$. The mean level of $PIR_{i,j}^{(t_1:t_2)}$ (green dash-dotted line) associated with its standard deviation (purple dash-dotted lines) are also plotted. Panel (b): the results for $|C^{(\cdot)}| = 3$ with $n = 3^3 = 27$ possible contact sequences.

*2) Response surface of the mean and variance of $PIR_{i,j}^{(t_1:t_2)}$*

Measure of sensitivity of potential infection risk $PIR_{i,j}^{(t_1:t_2)}$ of the HCP $j$ at the place $i$ over time $(t_1:t_2)$ was investigated. We denote the mean level and the variance of $PIR_{i,j}^{(t_1:t_2)}$ of all sequences given the number of close contacts $|C^{(\cdot)}|$ as $E[PIR_{i,j}^{(\cdot)}]$ and $Var[PIR_{i,j}^{(\cdot)}]$, respectively. Next, we defined two levels of $p_{X(m),k(m)\to j}^{(\cdot)}$: $P_{low} \in (0,0.5]$ and $P_{high} = P_{low} + 0.3$, and derived the response surfaces of the $E\left[PIR_{i,j}^{(\cdot)}\right]$ and $Var\left[PIR_{i,j}^{(\cdot)}\right]$ with respect to two inputs $P_{low}$ and $|C^{(\cdot)}|$. As shown in Fig. 3, the response surface of $E\left[PIR_{i,j}^{(\cdot)}\right]$ showed that a high probability of successful viral transmission $p_{X(m),k(m)\to j}^{(\cdot)}$ will result in an extremely high value of $E\left[PIR_{i,j}^{(\cdot)}\right]$, e.g., $E\left[PIR_{i,j}^{(\cdot)}\right] = 0.8336$ when $|C^{(\cdot)}|$ is only 3, $P_{low} = 0.3$, and $P_{high} = 0.7$.

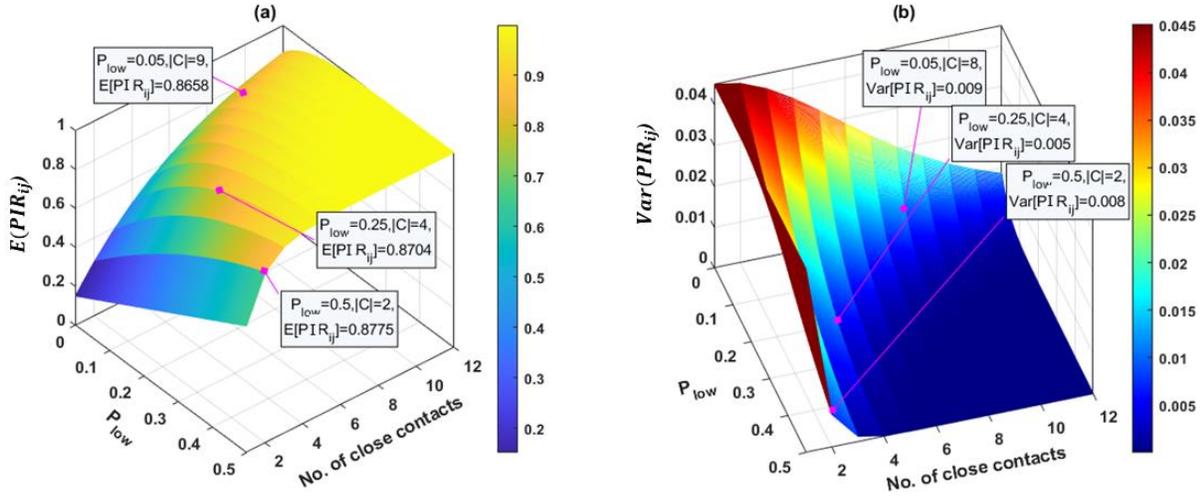

**Fig. 3.** Response surfaces of $E\left[PIR_{i,j}^{(t_1:t_2)}\right]$ and $Var\left[PIR_{i,j}^{(t_1:t_2)}\right]$ with respect to two input variables: viral transmission probability and number of close contacts. (a): the response surface of $E\left[PIR_{i,j}^{(t_1:t_2)}\right]$ subject to the change of $P_{low}$ and total number of close contacts $|C^{(\cdot)}| \in [1,12]$. A data set was synthesized with two levels of $p_{X(m),k(m)\to j}^{(\cdot)}$: $P_{low} \in (0,0.5]$ and $P_{high} = P_{low} + 0.3$. where the expectation $E\left[PIR_{i,j}^{(t_1:t_2)}\right]$ is the mean level of $PIR_{i,j}^{(t_1:t_2)}$ of all possible contact sequences $C^{(\cdot)}$, which are the combinations of $P_{low}$ and $P_{high}$ in the sequence of length $|C^{(\cdot)}|$. Data tips at 3 values of $P_{low}$: 0.05, 0.2, 0.5 were created to indicate the cut-off values of $|C^{(\cdot)}|$ when $E\left[PIR_{i,j}^{(t_1:t_2)}\right]$ was significantly high. Similarly, (b) shows the response surface of $Var\left[PIR_{i,j}^{(t_1:t_2)}\right]$ of all possible sequences subject to the change of $P_{low}$ and $|C^{(\cdot)}|$. Three data tips at $P_{low} = \{0.05, 0.2, 0.5\}$ were included to show the threshold of $|C^{(\cdot)}|$ at which $Var\left[PIR_{i,j}^{(t_1:t_2)}\right]$ was sufficiently low.

### B. Model validation using COVID-19 case study

Data sets of HCPs with COVID-19 were used to validate the proposed model. Access to these data sources can be provided per requests or via the cited references. The validation was performed on three main components: the viral transmission probability model, the individual-level infection risk model, and the population-level risk model. The HCP's occupational infection risk to COVID-19, interim guidance regarding risk assessment and universal PPE policy issued by the CDC [41], and the risk factors for severe acute respiratory syndrome coronavirus (SARS-CoV-2) transmission in hospital settings from previous studies were also included to develop the model for the case study.

*1) Contributing factors associated with nosocomial COVID-19 infection in healthcare workers*

The major factors resulting in high risk for HCPs are 1) exposure to COVID-19 patients without using appropriate PPE, 2) involvement in aerosol-generating procedures and the interventions performed by physicians or nurses, and 3) contact with patients and colleagues during the incubation period. Many studies suggested that there is a significant association between PPE use and infection risk and that masks are the most consistent contributing measure to reduce the risk. A similar association was observed for other PPE, such as gowns, gloves, and eye protection. Other exposures and treatment practices (e.g., intubation involvement, patient care, or having contact with secretions) were found to link with increased infection risk for HCPs [28, 29]. Finally, given the implementation of a universal PPE policy, the high risk of infection among HCP also arises from contacting asymptomatic patients and colleagues who are in the early phase of viral infections [19]. The risk factors for SARS-CoV-2 transmission in hospital settings identified by previous studies [17, 23] were also included to develop the model.

*2) Related work of HCP infection risk for COVID-19*

Different regression models, including logistic regression, log-binomial, and Poisson, were used with the defined risk measures to estimate the viral infection risk among HCP groups [18-20, 30-37]. Statistical survival analysis models were also used to estimate the HCP's risk of contracting SARS COV-2 viruses and the expected duration of time until viral infection occurs. Shah et al. [22] modeled hospital admission of healthcare workers with COVID-19 using Cox regression and conditional logistic regression. Long Nguyen et al. [21] assessed the COVID-19 infection risk among healthcare workers in contrast to the general community by examining the effect of PPE on risk. They also used Cox' proportional hazards model to calculate multivariate-adjusted hazard ratios (HRs) of a positive test. However, the major limitations of these models are: 1) the individual-specific characteristics, e.g., occupation, type of PPE used, experience level, and exposure duration to COVID-19 patients, are not considered [21, 22], and 2) the simple formalism of the models without time-varying stochastic transmissions oversimplifies the complex contagious mechanism of SARS COV-2.

*3) Data description*

Data collected from multiple sources (e.g., COVID-19 transmission databases, health surveys/questionaries, U.S.

Department of Labor databases, Cross-sectional study of UK-based healthcare workers) are illustrated in Table I.

TABLE I
SOURCES OF DATABASES INFORMATION INCLUDING SOURCE, NATION, UPDATED TIME, AND OWNER

| Data source | Nation | Updated time | Owner |
|---|---|---|---|
| Characteristics of HCP with COVID-19 [38] | US | July 16th, 2020 | U.S. CDC |
| COVID-19 transmission dynamics data [39] | Taiwan | Apr 2nd, 2020 | Taiwan CDC |
| California COVID-19 Health Surveys [40] | US | Sep 31st, 2020 | California COVID-19 Health Center |
| Texas Health Center COVID-19 Survey [41] | US | Oct 7th, 2020 | Texas Health Center |
| O*Net database [42] | US | Nov 16th, 2020 | U.S. Department of Labor |
| COVID-NET database [43] | US | Aug 28th, 2020 | U.S. CDC |
| Texas COVID-19 Data [44] | US | Apr 29th, 2021 | Texas Department of State Health Services |
| Cross-sectional observational study of UK-based HCP [45] | UK | May 25th, 2020 | The authors |

*4) Model variable selection*

Variables from recent findings of SARS-CoV-2 as introduced in Sub-section III.B.1, were used to select the features. The validation was performed on three main components: the viral transmission probability model, the individual-level infection risk model, and the population-level risk model. Regarding the viral transmission probability model, we included the following covariates in the model: $Age$, $Cancer$, $Resp$, $Obes$, $Smoker$, $Allied\_prof$, $Dental\_staff$, $Doctor$, $Pub\_trans$, $C\_contact$, $AGP$, $PPE\_train$, $Lacked\_PPE$, $Cont\_(wo\_PPE)$, and $Imp\_PPE$. These are significant factors suggested by the original cross-sectional study [45]. The description of these variables is summarized in Table 1S in the Supplementary material. To validate the individual-level infection risk model, the U.S. Department of Labor O*Net database was employed to quantify the risk score for healthcare-related occupations, where virus exposure time and duration and working environment were considered. For the population-level risk model, the PPE sufficiency level, regional infection risk and the hospitalization data of HCP were selected to estimate population-level infection risk in California and Texas medical centers [40, 41] and implement a surrogate method for model validation. The description of these variables is summarized in Table 1S in the Supplementary material.

*5) Model validation of viral transmission probability estimation using multivariate logistic regression*

To validate the logistic regression introduced in Sub-section II.A., we considered different protective and risk factors for COVID-19 in the data set of UK-based healthcare workers [45] and modelled the association between these covariates and the COVID-19 infection status using multivariable logistic regression. The data set provides 6263 responses in which a composite outcome was present in 1,806 (29.4%) HCP, of whom 49 (0.8%) were admitted to hospitals, 459 (7.5%) were tested positive for SARS-CoV-2, and 1,776 (28.9%) were self-isolated. The covariates included in the model were reported in Sub-section III.B.4. The estimated coefficients and their significance are shown in Table II.

TABLE II
ESTIMATED COEFFICIENT OF MULTIVARIATE LOGISTIC REGRESSION OF INFECTION RISK MODELING

| Variables | Coefficient estimates | SE | p-value |
|---|---|---|---|
| Intercept | -0.5953 | 0.1497 | 6.98e-05*** |
| $Age$ | -0.0120 | 0.0028 | 1.77e-05*** |
| $Cancer$ | 0.5296 | 0.2407 | 0.0277* |
| $Resp$ | 0.2020 | 0.0947 | 0.0328* |
| $Obes$ | 0.3055 | 0.0872 | 0.0004*** |
| $Smoker$ | -0.2490 | 0.1053 | 0.0180* |
| $Doctor$ | 0.1514 | 0.0662 | 0.0222* |
| $Allied\_prof$ | -0.2282 | 0.0852 | 0.0074** |
| $Dental\_staff$ | -0.7018 | 0.2113 | 0.0008*** |
| $Pub\_trans$ | 0.2728 | 0.0693 | 8.31e-05*** |
| $C\_contact$ | 0.2949 | 0.0724 | 4.63e-05*** |
| $AGP$ | -0.2201 | 0.0663 | 0.0009*** |
| $PPE\_train$ | -0.1708 | 0.0666 | 0.0104* |
| $Lacked\_PPE$ | 0.3237 | 0.0776 | 3.03e-05*** |
| $cont\_wo\_PPE$ | 0.3261 | 0.0768 | 2.21e-05*** |
| $Imp\_PPE$ | -0.2070 | 0.0865 | 0.0166* |

Significance codes: $p \approx 0$ '***', $p < 0.001$ '**', $p < 0.01$ '*', AIC: 7317.7

The model goodness-of-fit was further assessed by the Akaike information criterion (AIC) and 10-fold cross validation. The AIC value for the above model was 7317.70 and that for the null model was 7449.75. The 10-fold cross validation accuracy was calculated to be 78.23%, which showed that the performance on test data was relatively good.

*6) Model validation of the individual-level infection risk*

To validate to infection risk model at the individual level, six occupations were considered using the U.S. Department of Labor O*Net database [42]. We also introduced a new variable called occupational-specific risk score denoted as $ORS$ to account for the differences in infection risk among different occupations. The score was computed as:

$$ORS = \frac{(CO + PP + EI)}{3\phi} \times \frac{N_{hours}}{\max\{N_{hours}\}} \qquad (10)$$

where $\max\{N_{hours}\}$ is the maximum working hours per week of 6 occupations, and $\phi$ is the scaling parameter. The description of those variables $CO, PP, EI$, and $N_{hours}$ are summarized in Table 1S in the Supplementary material. Because of the limited longitudinal data, our strategy was to validate the individual infection risk model using hypothesized scenarios of different occupational settings. Particularly, we made four main assumptions: 1) the individual-risk is the same for every individual working under the same conditions (e.g., same occupation), 2) all patients are confirmed cases, i.e., there is only one compartment $IC$, 3) the probabilities of viral transmission from all patients are the same for each occupation, and 4) the probability of viral transmission estimate for confirmed infectious patients, denoted as $\hat{p}_{IC}^{(t_1:t_2)}$, is equal to $ORS/\max\{ORS\}$, where $\max\{ORS\}$ is the maximum $ORS$ score among 6 occupations, which guarantees $0 \leq p_{IC}^{(t_1:t_2)} \leq 1$. Consequently, (3) is reduced to:

$$PIR_{i,j}^{(t_1:t_2)} = \sum_{m=1}^{|C^{(\cdot)}|} \left(1 - \hat{p}_{IC}^{(t_1:t_2)}\right)^{m-1} \hat{p}_{IC}^{(t_1:t_2)} \quad (11)$$

Lastly, the total number of contacts $|C^{(\cdot)}|$ was fixed to be 5 and the value $\phi$ was set to 20. Next, the risk was estimated using (11), and the results are summarized in Table III.

TABLE III
ESTIMATED INDIVIDUAL-LEVEL INFECTION RISK FOR SIX DIFFERENT OCCUPATIONAL SETTING

| Occupations | ORS | $\hat{p}_{IC}^{(t_1:t_2)}$ | $PIR_{i,j}^{(t_1:t_2)}$ |
|---|---|---|---|
| Registered Nurses | 95.67 | 0.05 | 0.2262 |
| Personal Care Aides | 48.54 | 0.0254 | 0.1206 |
| Nursing Assistants | 59.08 | 0.0309 | 0.1451 |
| Medical Assistants | 89 | 0.0465 | 0.2119 |
| Licensed Nurses | 52.94 | 0.0277 | 0.1309 |
| Respiratory Therapists | 64.47 | 0.0337 | 0.1575 |

The results of the individual-level model indicated a strong positive association between the estimated risk $PIR_{i,j}^{(\cdot)}$ and virus transmission probability $p_{IC}^{(\cdot)}$, in which the top three occupations that have highest risk were registered nurses, medical assistants, and respiratory therapists. Their associated $PIR_{i,j}^{(\cdot)}$ values were 0.2262, 0.2119, and 0.1575 respectively, which were relatively high when $|C^{(\cdot)}| = 5$.

*7) Model validation of the population-level infection risk*

The population-level infection risk was validated based on the total of confirmed COVID-19 cases of HCP reported to the CDC. The number of positive COVID-19 cases of HCP in the US up to April 9, 2020, is presented in Fig. 4.

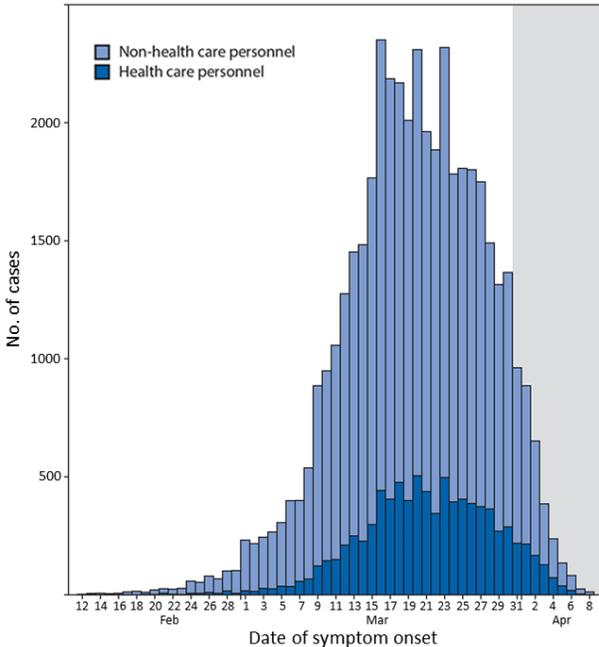

**Fig. 4.** Daily number of laboratory-confirmed positive COVID-19 cases by date of symptom onset of health care personnel and non-health care personnel (N = 43968) in the US from February 12 to April 9, 2020 [38].

According to Fig. 4, there was a strong association between the number of positive cases among non-HCP and the number of cases among HCP by date of symptom onset. In addition, the risk of infection among HCP was closely related to the total number of positive tests among HCP and the patient loads that HCP needed to handle. For population-level, we used the following selected features: $SOH_{time}$, $CS$, $PPE_{SL}$, $ORS$. The description of those is elaborated in Table 1S in the Supplementary material. Based on (8), population-level risk estimation was reduced to a regressive equation with equal weights assigned to each variable as:

$$\widehat{PIR}_i = \begin{pmatrix} E_{SOH_{time}}\left[PIR_{i,j}^{(\cdot)}\right] + E_{CS}\left[PIR_{i,j}^{(\cdot)}\right] + \\ E_{PPE_{SL}}\left[PIR_{i,j}^{(\cdot)}\right] + E_{ORS}\left[PIR_{i,j}^{(\cdot)}\right] \end{pmatrix}/4 \quad (12)$$

where $E_X\left[PIR_i^{(\cdot)}\right]$ is the expected value of $PIR_i^{(\cdot)}$ over the distribution of the variable $X$ and $Value(X)$ is the value set of $X$, $E_X\left[PIR_i^{(\cdot)}\right]$ is estimated as:

$$E_X\left[PIR_i^{(\cdot)}\right] = \sum_{x \in Value(X)} P(X = x) E\left[PIR_i^{(\cdot)} \big| X = x\right] \quad (13)$$

The population-level infection risk model was validated using the COVID-19 data from health centers in Texas, California and other relevant sources as presented in Subsection III.B.3 and Table I. The accessible HCP COVID-19 data of Texas and California were PPE sufficiency level, the total number of hospitalizations, and the percentage of ICU beds available. So, we assumed the distributions and the expected value of $PER_i^{(\cdot)}$ over the other variables to be the same for both states. The expected values of $PER_i^{(\cdot)}$ was computed using (13) (see Table IV).

TABLE IV
ESTIMATED VALUE AND DISTRIBUTION OF THE SELECTED FEATURES USED IN TWO CASE STUDIES TO ESTIMATE THE INFECTION RISK IN TEXAS AND CALIFORNIA

| Features | Texas | California |
|---|---|---|
| Time from symptom onset to hospitalization | The distributions of $SOH_{time}$ and $CS$ are estimated from [38, 39]. $P(SOH_{time} < 0) = 0.34, P(SOH_{time} \in [0,3]) = 0.21$ $P(SOH_{time} \in [4,5]) = 0.05$, $P(SOH_{time} \in [6,7]) = 0.02$, $P(SOH_{time} \in [8,9]) = 0.16, P(SOH_{time} > 9) = 0.21$ $E_{SOH_{time}}\left[PIR_i^{(\cdot)}\right] = 5.99 \times 10^{-3}$ | |
| Clinical severity of patients | $P(CS = \text{"Severe pneumonia"}) = 0.01$, $P(CS = \text{"ARDS/Sepsis"}) = 0.01$ $P(CS = \text{"Asymptomatic"}) = 0.08$, $E_{CS}\left[PIR_i^{(\cdot)}\right] = 3.89 \times 10^{-3}$ | |
| PPE sufficiency level | PPE sufficiency levels were averaged to estimate $E_{PPE_{SL}}\left[PIR_{i,j}^{(\cdot)}\right] = 0.0065$ | Similar to the calculation for Texas, $E_{PPE_{SL}}\left[PIR_{i,j}^{(\cdot)}\right]$ was estimated to be 0.744 |
| ORS | $E_{ORS}\left[PIR_i^{(\cdot)}\right]$ was estimated to be the average over of $PIR_{i,j}^{(t_1:t_2)}$ over all occupations at 0.0173 | |
| Estimated $\widehat{PIR}_i$ | $\widehat{PIR}_{Texas} = 0.0084$ | $\widehat{PIR}_{California} = 0.0132$ |

In Table IV, $E_{PPE_{SL}}\left[PIR_i^{(\cdot)}\right]$ was estimated using the PPE lacking information in health centers in Texas and HCP surveys in California. The value of $E_{ORS}\left[PIR_i^{(\cdot)}\right]$ was estimated by averaging the values of $PIR_{i,j}^{(\cdot)}$ over all occupations. The estimated $\widehat{PIR}_i$ values for Texas and California were 0.0084 and 0.0132, respectively.

## IV. DISCUSSION

In our sensitivity analysis, we focused only on two key variables, namely viral transmission probability and the number of close contacts between HCP and patients. Specifically, the sensitivity of the infection risk to those input variables was measured by the amount of variance caused by changing the inputs. We divided our analysis into two parts: 1) the measure of sensitivity of $PIR_{i,j}^{(\cdot)}$ to $p_{X(m),k(m)\to j}^{(\cdot)}$ and close contact sequence, and 2) response surface of the mean and variance of $PIR_{i,j}^{(t_1:t_2)}$ to $|C^{(\cdot)}|$ and $p_{X(m),k(m)\to j}^{(\cdot)}$. The results of the sensitivity analysis revealed that the output $PIR_{i,j}^{(\cdot)}$ will be significantly increased when the viral transmission probability $p_{X(m),k(m)\to j}^{(\cdot)}$ and the number of close contacts become heighten ,. In addition, the results in the second part indicated that $E[PIR_{i,j}^{(\cdot)}]$ quickly converged to one as $|C^{(\cdot)}| \to \infty$, and the convergence rate was faster if $P_{low}$ took higher values. Based on the response surface of $Var[PIR_{i,j}^{(\cdot)}]$, higher values of $P_{low}$ and $|C^{(\cdot)}|$ will lead to a lower value of $Var[PIR_{i,j}^{(\cdot)}]$; however, the effect of $P_{low}$ is more significant than that of $|C^{(\cdot)}|$. The value of $Var[PIR_{i,j}^{(\cdot)}] \to 0$ as $|C^{(\cdot)}| \to \infty$ and dropped to nearly 0 after only four close contacts when $P_{low} = 0.5$.

After performing the sensitivity analysis, the logistic regression for estimating viral transmission probability $\hat{p}_{X,r\to j}^{(t_1:t_2)}$ was validated using the cross-sectional observational study of UK-based healthcare workers. Based on the coefficient estimates of the variables in the built multivariate logistic regression model, $Age$, $Smoker$, $Allied\_prof$, $Dental\_staff$, $AGP$, $PPE\_train$, $Imp\_PPE$ were the protective factors, whereas the risk factors were $Cancer$, $Resp$, $Obes$, $Doctor$, $Lacked\_PPE$, $cont\_wo\_PPE$, $Pub\_trans$, $C\_contact$. Surprisingly, advanced age, being a smoker or ex-smoker within one year, and having regular exposure to aerosol-generating procedures performed on COVID-19 patients decreased the infection risk. This result seems counter-intuitive at first, but they are confounders because it was shown that HCP working directly with suspected or confirmed COVID-19 patients tended to be more cautious and self-aware in clinical environments[46]. Therefore, they had sufficient self-protection and took containment measures; however, healthcare workers in non-communicable viral disease departments, who were potentially exposed to contagious viruses, did not have sufficient training on how to use PPE and deal with infectious diseases and lack of access to PPE and isolation equipment [47]. However, the model has several limitations. First, because we did not have access to information on HCP contact with patients and coworkers, we assumed the estimated viral transmission probability as a measure averaged over all individuals. Second, the data were gathered using surveys and questionnaires, which are subject to selection and recall bias. Third, the use of a composite outcome (including HCP with COVID-19 symptoms, HCP being exposed to risk factors, and lab-confirmed HCP infections) may have resulted in overestimation or underestimation of the infection risk.

We validated the individual-level infection risk model, implemented the model using the two-parameter regressive equation, and estimated the individual risk for six occupations. The results highly depend on the pre-defined parameters, which can be estimated in healthcare settings when data are available. It was shown that healthcare workers and nurses are frequently in close contact with COVID-19 patients, which therefore increases the risk for acquiring SARS-CoV-2 virus [48]. Because HCP can acquire infection through various pathways apart from direct patient care, such as exposure to colleagues, family members, or people in the community, the time-varying risk estimation in the model can provide informed decisions for screening HCP for COVID-19 before workplace entry. The individual risk model can be improved and more specific to better model the transmission dynamics, e.g., a model that incorporates the quantification of indoor airborne infection risks using a probabilistic framework [49].

For model validation at the population level, we considered two case studies to estimate the risk of infection of HCP in Texas and California states. Both states have a high number of lab-confirmed SARS-CoV-2 patients. The average number of hospitalizations in Texas and California were 16843 cases/day and 4219 cases/day, respectively. However, the infection risk in Texas was 0.0084 which was lower than the risk in California (0.0132). This was mainly due to the difference in patient load for each HCP per day and the two states' PPE sufficiency level. From Table IV, the average PPE sufficiency level in California was only 0.744 as opposed to 0.9355 in Texas, and the average percentage of ICU beds available per 100,100 people in Texas was significantly higher than that in California, which implies heavier patient loads in California. The model also made some important assumptions: 1) close contacts with COVID-19 patients are independent and there is no viral transmission among HCP, and 2) protective/risk factors are well-defined and sufficient to estimate the risk of infection.

## V. CONCLUSION AND FUTURE WORK

The paper proposed a time-variant infection risk analysis model to characterize the dynamic of the disease infection risk in HCP over time and an individual-specific and domain-knowledge driven infection risk to quantify the complexities of HCP's risk of CVDs in healthcare settings. The infection risk analysis model for HCP was estimated at both individual and population levels. The individual-level risk model was built based on the population grouping concept of the well-established epidemiological SEIR model with the consideration of the time-varying confounders to capture the dynamical contagious disease transmission mechanism. At the population-level, three subsets of features were constructed and represented by a Bayesian network, from which the probability of viral transmission from patients to HCP was estimated. To validate our methods, we have incorporated the data from multiple data sources from the US, the UK, and Taiwan for the COVID-19 case study, which contains the information about potential factors that affect COVID-19 transmission mechanism; and the domain knowledge of similar contagious diseases such as SARS or MERS from the relevant studies to estimate the risk of COVID-19 infection of HCP. For individual-level risk estimation, the model was founded on the SEIR compartmental model and developed for the occupational-specific and individualized infection risk model. As a result, the model can capture accurately the infection risk varying over time under the control of those individual time-varying confounders, and it is also able to account for the intrinsic stochastic transmission mechanisms. At the population level, the Bayesian network formalism can accommodate the limited data scenario, and it can update the parameters when more data are available. The results from two case studies are interpretable at the population level, which showed infection risk in California is higher than in Texas because of the heavier patient loadings and shortage of PPE. The major limitations of the CDC's interim guideline for risk assessment, which is inadequate in quantifying the risk of infection in an individualized HCP, have been addressed by our model. The model would significantly endorse the PPE allocation and safety plans for HCP and enhance the crisis-level staffing strategies in facilities with the staffing shortage. Longitudinal experimental designs are required to collect more COVID-19 data among HCP to validate the proposed model properly. Future work would involve: 1) model assumption validation when more data are available and sufficient, 2) model modification and reformulation if the assumptions are violated (e.g., independence assumption and new vaccinated population), and 3) validating the model with the other related case studies of communicable viral diseases.

## ACKNOWLEDGMENT

We would like to thank doctor Arveity R. Setty for assistance with the problem formulation in the hospital settings, and Prof. Om P. Yadav for the comments and ideas that greatly improved the manuscript.

## REFERENCES

1. English, K.M., et al., *Contact among healthcare workers in the hospital setting: developing the evidence base for innovative approaches to infection control.* BMC infectious diseases, 2018. **18**(1): p. 1-12.
2. Gastmeier, P., et al., *How outbreaks can contribute to prevention of nosocomial infection: analysis of 1,022 outbreaks.* Infection Control & Hospital Epidemiology, 2005. **26**(4): p. 357-361.
3. Gastmeier, P. and R.-P. Vonberg, *Outbreaks of nosocomial infections: lessons learned and perspectives.* Current opinion in infectious diseases, 2008. **21**(4): p. 357-361.
4. Iversen, K., et al., *Risk of COVID-19 in health-care workers in Denmark: an observational cohort study.* The Lancet Infectious Diseases, 2020. **20**(12): p. 1401-1408.
5. Mutambudzi, M., et al., *Occupation and risk of severe COVID-19: prospective cohort study of 120 075 UK Biobank participants.* Occupational and Environmental Medicine, 2020.
6. McDougal, A.N., et al., *Outbreak of coronavirus disease 2019 (COVID-19) among operating room staff of a tertiary referral center: An epidemiologic and environmental investigation.* Infection Control & Hospital Epidemiology, 2021: p. 1-7.
7. Khatib, A.N., et al., *Navigating the risks of flying during COVID-19: a review for safe air travel.* Journal of travel medicine, 2020. **27**(8): p. taaa212.
8. Cooper, B.S., *Confronting models with data.* Journal of Hospital Infection, 2007. **65**: p. 88-92.
9. Grundmann, H. and B. Hellriegel, *Mathematical modelling: a tool for hospital infection control.* The Lancet infectious diseases, 2006. **6**(1): p. 39-45.
10. Brauer, F., *Compartmental models in epidemiology*, in *Mathematical epidemiology*. 2008, Springer. p. 19-79.
11. Di Stefano, B., H. Fuks, and A.T. Lawniczak. *Object-oriented implementation of CA/LGCA modelling applied to the spread of epidemics.* in *2000 Canadian Conference on Electrical and Computer Engineering. Conference Proceedings. Navigating to a New Era (Cat. No. 00TH8492).* 2000. IEEE.
12. Sirakoulis, G.C., I. Karafyllidis, and A. Thanailakis, *A cellular automaton model for the effects of population movement and vaccination on epidemic propagation.* Ecological Modelling, 2000. **133**(3): p. 209-223.
13. Zhen, J. and L. Quan-Xing, *A cellular automata model of epidemics of a heterogeneous susceptibility.* Chinese Physics, 2006. **15**(6): p. 1248.
14. Casalicchio, E., E. Galli, and S. Tucci, *Agent-based modelling of interdependent critical infrastructures.* International Journal of System of Systems Engineering, 2010. **2**(1): p. 60-75.
15. Perez, L. and S. Dragicevic, *An agent-based approach for modeling dynamics of contagious disease spread.* International journal of health geographics, 2009. **8**(1): p. 1-17.
16. Voirin, N., et al., *A multiplicative hazard regression model to assess the risk of disease transmission at hospital during community epidemics.* BMC medical research methodology, 2011. **11**(1): p. 1-8.
17. Chu, D.K., et al., *Physical distancing, face masks, and eye protection to prevent person-to-person transmission of SARS-CoV-2 and COVID-19: a systematic review and meta-analysis.* The lancet, 2020. **395**(10242): p. 1973-1987.
18. Wang, Q., et al., *Epidemiological characteristics of COVID-19 in medical staff members of neurosurgery departments in Hubei province: a multicentre descriptive study.* medRxiv, 2020.
19. Eyre, D.W., et al., *Differential occupational risks to healthcare workers from SARS-CoV-2 observed during a prospective observational study.* Elife, 2020. **9**: p. e60675.
20. Ki, H.K., et al., *Risk of transmission via medical employees and importance of routine infection-prevention policy in a nosocomial outbreak of Middle East respiratory syndrome (MERS): a descriptive analysis from a tertiary care hospital in South Korea.* BMC pulmonary medicine, 2019. **19**(1): p. 1-12.
21. Nguyen, L.H., et al., *Risk of COVID-19 among front-line health-care workers and the general community: a prospective cohort study.* The Lancet Public Health, 2020. **5**(9): p. e475-e483.
22. Shah, A.S., et al., *Risk of hospital admission with coronavirus disease 2019 in healthcare workers and their households: nationwide linkage cohort study.* bmj, 2020. **371**.
23. Friedman, N., D. Geiger, and M. Goldszmidt, *Bayesian network classifiers.* Machine learning, 1997. **29**(2-3): p. 131-163.
24. Chou, R., et al., *Epidemiology of and risk factors for coronavirus infection in health care workers: a living rapid review.* Annals of internal medicine, 2020. **173**(2): p. 120-136.
25. Kallenberg, O., *Random measures, theory and applications*. Vol. 1. 2017: Springer.
26. Huynh, P.K., et al., *Probabilistic domain-knowledge modeling of disorder pathogenesis for dynamics forecasting of acute onset.* Artificial Intelligence in Medicine, 2021. **115**: p. 102056.
27. Jensen, F.V., *Bayesian networks.* Wiley Interdisciplinary Reviews: Computational Statistics, 2009. **1**(3): p. 307-315.
28. Chen, Q., A. Allot, and Z. Lu, *Keep up with the latest coronavirus research.* Natur, 2020. **579**(7798): p. 193-193.
29. Raboud, J., et al., *Risk factors for SARS transmission from patients requiring intubation: a multicentre investigation in Toronto, Canada.* PLoS One, 2010. **5**(5): p. e10717.
30. Caputo, K.M., et al., *Intubation of SARS patients: infection and perspectives of healthcare workers.* Canadian Journal of Anesthesia, 2006. **53**(2): p. 122.
31. Chen, W.-Q., et al., *Which preventive measures might protect health care workers from SARS?* BMC Public Health, 2009. **9**(1): p. 1-8.
32. Le Dang Ha, S.A.B., et al., *Lack of SARS transmission among public hospital workers, Vietnam.* Emerging infectious diseases, 2004. **10**(2): p. 265.
33. Alraddadi, B.M., et al., *Risk factors for Middle East respiratory syndrome coronavirus infection among healthcare personnel.* Emerging infectious diseases, 2016. **22**(11): p. 1915.
34. Hall, A.J., et al., *Health care worker contact with MERS patient, Saudi Arabia.* Emerging infectious diseases, 2014. **20**(12): p. 2148.
35. Bai, Y., et al., *SARS-CoV-2 infection in health care workers: a retrospective analysis and a model study.* medRxiv, 2020.
36. Heinzerling, A., et al., *Transmission of COVID-19 to health care personnel during exposures to a hospitalized patient—Solano County, California, February 2020.* 2020.

37. Mutambudzi, M., et al., *Occupation and risk of severe COVID-19: prospective cohort study of 120 075 UK Biobank participants.* Occupational and Environmental Medicine, 2021. **78**(5): p. 307-314.
38. COVID, T.C., *Characteristics of Health Care Personnel with COVID-19-United States, February 12-April 9, 2020.* https://www.cdc.gov/mmwr/volumes/69/wr/pdfs/mm6915e6-H.pdf, 2020.
39. Cheng, H.-Y., et al., *Contact tracing assessment of COVID-19 transmission dynamics in Taiwan and risk at different exposure periods before and after symptom onset.* JAMA internal medicine, 2020.
40. *California Health Care Foundation, California COVID-19 Health Surveys: Data and Charts.* April 1, 2020; Available from: https://www.chcf.org/project/california-covid-19-health-surveys/#physician-survey.
41. *Health Resources & Services Adminstration, Texas Health Center COVID-19 Survey Summary Report.* Oct 7th, 2020; Available from: https://bphc.hrsa.gov/emergency-response/coronavirus-health-center-data/tx.
42. *O*Net database.* Nov 16th, 2020; Available from: https://www.onetonline.org/.
43. Garg, S., *Hospitalization rates and characteristics of patients hospitalized with laboratory-confirmed coronavirus disease 2019—COVID-NET, 14 States, March 1–30, 2020.* MMWR. Morbidity and mortality weekly report, 2020. **69**.
44. *Texas COVID-19 Data.* Apr 29th, 2021; Available from: https://dshs.texas.gov/coronavirus/additionaldata.aspx.
45. Kua, J., et al., *healthcareCOVID: a national cross-sectional observational study identifying risk factors for developing suspected or confirmed COVID-19 in UK healthcare workers.* PeerJ, 2021. **9**: p. e10891.
46. Du, Q., et al., *Nosocomial infection of COVID-19: A new challenge for healthcare professionals.* International Journal of Molecular Medicine, 2021. **47**(4): p. 1-1.
47. McMichael, T.M., et al., *Epidemiology of Covid-19 in a long-term care facility in King County, Washington.* New England Journal of Medicine, 2020. **382**(21): p. 2005-2011.
48. Hughes, M.M., et al., *Update: characteristics of health care personnel with COVID-19—United States, February 12–July 16, 2020.* Morbidity and Mortality Weekly Report, 2020. **69**(38): p. 1364.
49. Liao, C.M., C.F. Chang, and H.M. Liang, *A probabilistic transmission dynamic model to assess indoor airborne infection risks.* Risk Analysis: An International Journal, 2005. **25**(5): p. 1097-1107.

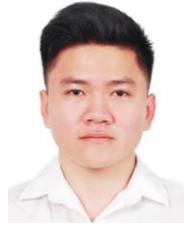

**Phat K. Huynh** is currently a PhD student of the Industrial and Manufacturing Engineering Department at North Dakota State University. He earned a bachelor's degree in biomedical engineering, International University - VNU-HCM in 2018, Viet Nam. His work mainly focuses on wearable devices, nonlinear dynamical systems, probabilistic statistical models, machine learning, and predictive analytics in healthcare

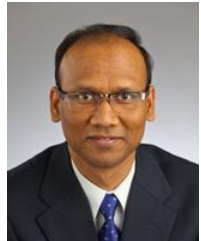

**Arveity Setty** is a Sleep Medicine Specialist in Fargo, ND and has over 9 years of experience in his field. He graduated from Michigan State University, College of Human Medicine Medical School in 2012. He is affiliated with Sanford Health. He serves as a Clinical Associate Professor at the University of North Dakota School of Medicine & Health Services. Dr. Arveity Setty practices pediatric sleep medicine. He specializes in sleep studies and actigraphy.

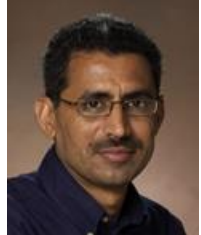

**Om P. Yadav** is currently a Professor and Chair of the Department of Industrial and Systems Engineering at North Carolina A&T State University. He has authored or coauthored more than 150 research articles in different areas namely reliability, risk assessment, and design optimization. His research interests include reliability modeling and analysis, risk assessment, design optimization and robust design, and manufacturing systems analysis.

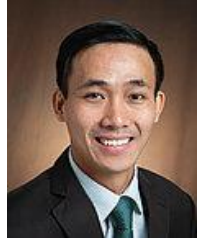

**Trung Q. Le** is an Assistant Professor of the Industrial and Manufacturing Engineering Department at North Dakota State University. He was an Assistant Professor and Associate Department Head for Research Affairs in Biomedical Engineering Department at International University - VNU-HCM, Vietnam. His research focuses in 3 main directions: 1) Data-driven and Sensor-based Modeling, 2) Medical Device Manufacturing and Bio-signal Processing, and 3) Predictive Analytics for Personalized Healthcare.